\documentclass[prd,reprint,nofootinbib,amsmath,amssymb,aps,floatfix]{revtex4-1}
\usepackage{amssymb,amsmath,mathrsfs, graphicx,color}
\usepackage[linktoc=page]{hyperref}
\usepackage{subfigure}
\usepackage{amsfonts}
\usepackage[utf8]{inputenc} 
\usepackage[normalem]{ulem}

\newcommand{\beq}{\begin{equation}}
\newcommand{\eeq}[1]{\label{#1}\end{equation}}
\newcommand{\bea}{\begin{eqnarray}}
\newcommand{\eea}[1]{\label{#1}\end{eqnarray}}

\newcommand{\scri}{\mathscr{I}}

\DeclareMathOperator{\Vol}{Vol}

\begin{document}

\title{Asymptotic Charges Cannot Be Measured in Finite Time}
\author{Raphael Bousso, Venkatesa Chandrasekaran, and Illan F. Halpern}%
 \email{bousso@lbl.gov, ven\_chandrasekaran@berkeley.edu, \\illan@berkeley.edu}
\affiliation{Center for Theoretical Physics and Department of Physics\\
University of California, Berkeley, CA 94720, USA 
}%
\affiliation{Lawrence Berkeley National Laboratory, Berkeley, CA 94720, USA}
\author{Aron C. Wall}%
 \email{aroncwall@gmail.com}
\affiliation{Varian Laboratory of Physics \\382 Via Pueblo Mall,
Stanford, CA 94305-4060}
\begin{abstract}
To study quantum gravity in asymptotically flat spacetimes, one would like to understand the algebra of observables at null infinity. Here we show that the Bondi mass cannot be observed in finite retarded time, and so is not contained in the algebra on any finite portion of ${\scri}^+$. This follows immediately from recently discovered asymptotic entropy bounds. We verify this explicitly, and we find that attempts to measure a conserved charge at arbitrarily large radius in fixed retarded time are thwarted by quantum fluctuations. We comment on the implications of our results to flat space holography and the BMS charges at $\scri^+$.  
\end{abstract}

\maketitle

\section{Communication Without Energy? \label{paradox}}

\begin{figure}[h]
\includegraphics[width=0.37\textwidth]{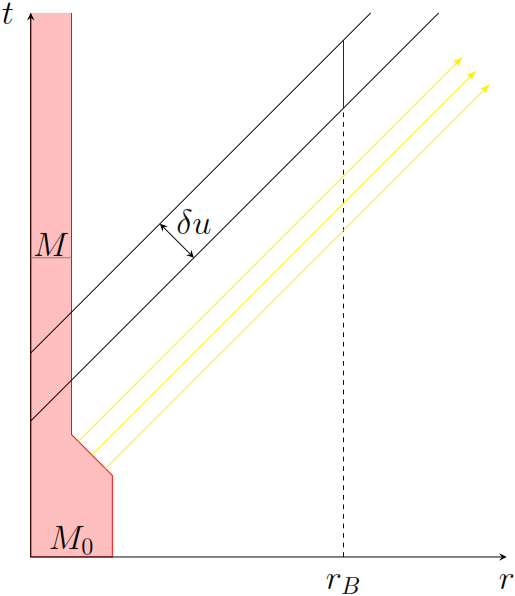} 
\caption{If distant observer Bob could measure the Bondi mass of Alice's planet, then Bob could receive information from Alice, without receiving energy. This would contradict recently proven bounds on distant communication channel capacities. In our example, Alice has radiated away some portion of her planet, but Bob does not intercept this radiation (yellow arrows). Instead, Bob later tries to measure how much mass is still left, in some fixed amount of time $\delta u$, at arbitrarily large radius $r_B$. We resolve the contradiction by showing that quantum fluctuations ruin Bob's measurement. The Bondi mass cannot be observed in finite time.}
\label{fig-alicebob}
\end{figure}

Alice would like to send Bob a message. Alice lives on a small, massive planet. Bob occupies a Dyson sphere of large radius $r_B$ and negligible mass, which surrounds Alice in an otherwise empty, asymptotically flat spacetime (see Fig.~\ref{fig-alicebob}). It would be simplest for Alice to send Bob a radio signal, or some gravitational waves. Unfortunately, their sleep schedules are out of sync, so that Bob would not be awake when Alice's signal arrives. Instead, they come up with an ingenious protocol, which makes it unnecessary for Bob to intercept any signal from Alice.

Their protocol is as follows. Long ago, before Bob traveled to the Dyson sphere, Alice told Bob the mass $M_0$ of her planet. She promised not to radiate any of it away until the agreed time when the message is to be sent. That fateful night, she radiates away a certain portion of the mass of her planet. The radiation passes through Bob's sphere while he sleeps, without interacting, and is lost forever. 

But when Bob wakes up, he measures the new Bondi mass $M$ of Alice's planet. This can be done at arbitrary distance, by measuring the surface integral that defines the Bondi mass (see Eqs.~(\ref{Bondi Expansion}) and (\ref{Bondi Mass}) below). 

Alice and Bob have agreed on a code, whereby the possible values of
$M$ are binned into discrete intervals, and each interval means a particular message. For example, suppose that Alice's planet has initial mass $M_0=10^{24}$ kg, and Bob is able to measure the final Bondi mass $M$ to a resolution of $1$ kg. Then Alice can choose from among $10^{24}$ messages. Upon measuring $M$, Bob gains an amount $\log 10^{24}$ of information, or about 80 bits. 

Alice and Bob believe that their scheme will work, given a sufficiently long but fixed, finite retarded time $\delta u$ for Bob to perform measurements after he wakes up, no matter how big the Dyson sphere is. That is, it should succeed in the limit as $r_B\to\infty$ at fixed retarded time $u\equiv t-r$ and fixed $\delta u$ (see Fig.~\ref{fig-alicebob}).

The restriction to fixed $u$ and $\delta u$ at arbitrarily large $r_B$ is very important to Bob, because he likes to finish all his work before his mid-afternoon nap. It is also important to many theorists, who wish to associate a Bondi mass (and other charges) to a ``cut,'' or cross-section, of future null infinity ${\scri}^+$, which lies at infinite $r$ and is parametrized by $u$. Of course, no measurement can be performed truly instantaneously, so Bob instead pursues the more modest goal of measuring the Bondi mass in some finite retarded time interval of length $\delta u$.

\begin{figure}[h] 
\includegraphics[width=0.25 \textwidth]{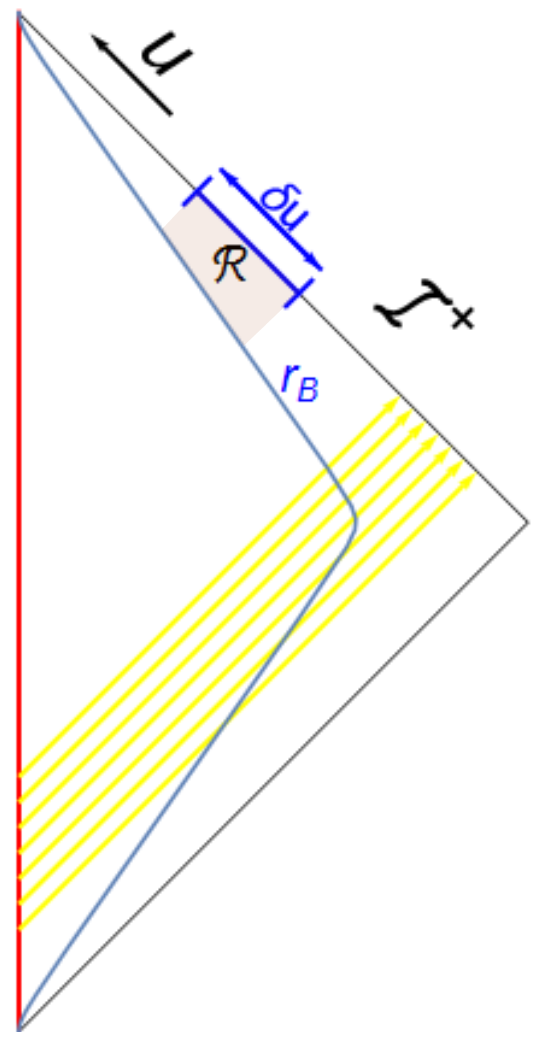}
\caption{ Penrose diagram of the process we consider. The red line represents Alice's worldline. The yellow arrows are the radiation emitted by Alice and reaching $\scri^+$ without interacting with Bob (blue worldline) whose detectors are only on for a retarded time interval $\delta u.$ 
}\label{fig-abpenrose}
\end{figure}

The formal definition of the Bondi mass is associated with a constant-$u$ cut of future null infinity, ${\scri}^+$ (see Fig.~\ref{fig-abpenrose}). To make contact with this definition, we consider the limit of a very large Dyson sphere, $r_B\to \infty$, at fixed retarded time $u_0$ in the metric
\begin{equation}
ds^2 = -\left(1 - \frac{2m_B}{r}\right)du^2 - 2 du\, dr + r^2 d\Omega^2 +\ldots \label{Bondi Expansion}
\end{equation}
The ellipsis indicates terms subleading in $1/r$ that we will not need. Here $m_B$ is the Bondi mass aspect. Its integral over a 2-sphere cut of $\mathcal{I}^+$ yields the Bondi mass:
\begin{align}
M = \frac{1}{4\pi}\int_{S^2}d^2\Omega \ m_B \label{Bondi Mass}
\end{align}
To claim that an asymptotic observer can measure the Bondi mass in finite time, is to claim that $M$ can be determined by measurements in a distant region ${\cal R}$ in Fig.~\ref{fig-abpenrose}. Here ${\cal R}$ is bounded on the inside by an arbitrarily large radius $r_B$, and in the past and future by the lightsheets $u=u_0\pm \frac{\delta u}{2}$.

However, if this protocol succeeded, we would have a paradox. Building on universal entropy bounds~\cite{BouCas14a,BouCas14b,BouFis15a,BouFis15b,Bou16,BouHal16}, it was recently shown that communication from Alice to Bob is constrained by a universal limit on the mutual information that can be achieved~\cite{Bou16b}. 

In the limit as $r_B\to\infty$, the amount of information that can be gained by Bob is of order $E \delta u$, where $E$ is the average energy of the signal that is actually received by his detectors. More precisely, the entropy in the detection region is bounded by the modular energy $K$ in the interval $\delta u$:
\begin{equation}
K =  \int d^2\Omega \int_{u_1(\Omega)}^{u_2(\Omega)}  du\, g(u)\, {\cal T}(u,\Omega)~.
\end{equation}
Here $\Omega$ is the angle on the sphere at ${\scri}^+$; ${\cal T} = \lim r^2 T_{uu} $ is the energy flux arriving on ${\scri}^+$ per unit angle and unit retarded time; and $g(u)$ is a positive definite function. (For a free field, $g(u) = \frac{(u_2-u)(u-u_1)}{u_2-u_1}$.)
But $K$ vanishes because ${\cal T}$ vanishes: Bob receives no energy at all. He missed the radiation Alice sent earlier, and by the time he measures the mass or charge, there is no radiative flux at all. The entropy is closely related to the Holevo quantity~\cite{Bou16b}, which bounds the mutual information between Alice and Bob. Hence, Bob cannot learn anything from Alice in this protocol.

In light of this contradiction, it is natural to go back and ask where the troublesome bound on communication~\cite{Bou16b} came from. It was obtained~\cite{Bou16,BouHal16} as a limit of the ``Quantum Bousso bound,'' which was proven for free field theories in~\cite{BouFis15a} and for interacting theories in~\cite{BouFis15b}. Ultimately, this entropy bound arose from the conjecture~\cite{CEB1,FMW} that the entropy in a region is bounded by the cross-sectional area loss along a {\em lightsheet} traversing the region, measured in Planck units. Here, the lightsheet is a family of parallel light-rays that pass through the asymptotic region. Radiation will focus such light-rays, and the area they span will contract by an amount that remains fixed in Planck units, as the location of the family is taken to infinite distance. The curvature due to the Schwarzschild metric of Alice's planet will also focus the light-rays (through a shear term), but it is easy to check that the resulting area loss goes to zero as the lightsheet is taken off to null infinity.

Thus, Alice and Bob's protocol must fail: it cannot be possible to extract information by measuring a conserved charge in fixed finite time at arbitrarily large distance. In this paper, we will show how it fails. We find that, in the limit as $r_B\to\infty$ at fixed $\delta u$, quantum fluctuations dominate and prevent Bob from measuring the conserved charge.\footnote{Astronomical determinations of mass are performed in the opposite limit, $\delta u\gg r_B$, and so are unconstrained by our analysis. For example, the mass of the Sun can be found by measuring the period of Earth and applying Kepler's Third Law. In such an experiment one has $r_B= 1$ A.U.\ $\approx 8$ min $\ll \delta u \sim 1$ year.}

This does not mean, of course, that it is impossible to measure a conserved charge at great distances. It just cannot be done in fixed finite time. As long as the duration of the measurement scales as an appropriate positive power of $r$, it is possible to determine the charge. But then the measurement cannot be associated with a finite neighborhood of a cut at future null infinity. Rather, the support of any successful measurement must approach (at least) a semi-infinite region of ${\scri}^+$ in the large $r$ limit. Similar comments apply to charges defined at spatial infinity, such as the ADM mass. They are defined by taking $r \to\infty$ at fixed $t$ rather than fixed $u$. Again the duration of the measurement must scale as a positive power of $r$ to control fluctuations.

\paragraph*{Outline} In Sec.~\ref{QED} we begin with warm-up problem: we consider charge fluctuations near future null infinity in massless QED. We turn to the gravitational case in Sec.~\ref{gravity}. An appendix contains details of our calculations.

\section{Bondi electric charge \label{QED}}

In standard QED, the charged particles are massive. Here we consider massless QED, as a closer analogue to the above thought-experiment where Alice uses a massless field (gravitons) to radiate away part of her planet's mass. Translated to the setting of massless QED, the paradox outlined above persists: Alice's planet now starts out with some nonzero charge $Q_0$, and Alice reduces this charge to $Q$ by emitting massless charged particles. The charged radiation crosses Bob's sphere while he sleeps, so when he later attempts to determine $Q$, he does so by measuring the radial electric field $E_r$ integrated over his Dyson sphere, and applying Gauss's law:
\begin{equation}
Q= r_B^2 \oint E_r(\Omega) d^2\Omega~,
\label{eq-q}
\end{equation}
where $\Omega$ is the solid angle on the sphere.

The fluctuation of the electric charge in some region, $\langle Q^2 \rangle$, can be computed by integrating the two-point function of the timelike component of the current density, $\langle j^0(x) j^0(y)\rangle$. Note that Bob does not attempt to measure $Q$ by integration of a charge density over a volume. Bob has access only to an asymptotic region, so naturally he would try to measure $Q$ by integrating the radial electric field over the boundary of the volume. But by Gauss's law, this is the same operator. Here we find it easier to evaluate its fluctuations using the volume form of the operator.

In any CFT, the two-point function is fixed by conformal invariance. In flat space the $U(1)$ current two-point function just takes the form \cite{Ginsparg:1988ui},
\begin{equation}
\langle j^0(x) j^0(y) \rangle = \kappa \frac{|\vec \Delta|^2 + (\Delta^0)^2}{\Delta^{8}}~, 
\label{eq-2j}
\end{equation}
where $\Delta=x-y$, and the constant $\kappa$ is theory dependent. For massless Dirac fermions, the current and the propagator are given by \cite{Peskin:1995ev}
\begin{eqnarray}
j^\mu & = & \bar \psi \gamma^\mu \psi~,\\ \label{eq:2pferm}
\langle \bar \psi(x) \psi(y) \rangle & = & -\frac{i}{2 \pi^2} \frac{\gamma_\mu (x^\mu - y^\mu)}{(x-y)^4}~,
\end{eqnarray}
which leads to $\kappa_{(\frac{1}{2})}=-\frac{1}{\pi^4}$. For comparison, in massless scalar QED one has\footnote{This is the leading order result.  Scalar QED is not really scale-invariant, due to the nontrivial renormalization group flow of the couplings.  Unlike a massless fermion field, $\phi$ can gain a mass by renormalization.  Even if one tunes the field to be massless, there will still be a logarithmic screening of the QED coupling constant as we flow to the IR.  However, since we find a power law divergence for $\langle Q^2 \rangle$ at leading order, it does not seem possible that this divergence can be removed by a logarithmic effect. Thus we expect our qualitative conclusions to be the same for massless scalar QED, as for the fermion.}
\begin{eqnarray}
j^\mu & = & i \left(\phi \partial^\mu \phi^* - \phi^* \partial^\mu \phi \right)~, \\
\langle \phi^*(x) \phi(y) \rangle & = & \frac{1}{4 \pi^2 (x-y)^{2}} , \label{eq:2pscalar}
\end{eqnarray}
which gives $\kappa_{(0)}=-\frac{1}{4 \pi^4}$.

In the 2-point functions~(\ref{eq:2pferm}) and (\ref{eq:2pscalar}), an $i \epsilon$ prescription must be specified. The choice 
\begin{equation}
\Delta^0\to \Delta^0 - i \epsilon
\end{equation}
allows for only non-negative energy states in the spectrum. In the complex $\Delta^0$-plane this corresponds to a contour prescription that cuts above both poles in Eq.~(\ref{eq-2j}). In what follows, this prescription will be implicit.

The total charge inside a spatial region $V$ at the time $t_B$ of Bob's measurement is
\begin{equation}
Q[V]= \int_V d^3 x \, j^0(x)~;
\end{equation}
but as an operator this would have divergent fluctuations. To obtain a well-defined operator, we smear over a finite time, 
\begin{equation}
Q= \int dt\,Q[V(t)]\, w(t)~.
\end{equation}
The weight function $w(t)$ is normalized so that $\int_{-\infty}^\infty w(t) dt=1$. It should peak in a finite time interval of characteristic size $\delta t$, centered on $t_B$; and it should fall off rapidly outside this interval. Our choice
\begin{equation}
w(t) = \frac{\delta t}{\pi} \frac{1}{(t-t_B)^2 + \delta t^2} \label{eq:w}~, 
\end{equation}
facilitates the application of contour integration methods. Any other choice with a fast enough fall off should lead to the same qualitative behavior.

For $V(t)$, we must choose the volume enclosed by Bob's Dyson sphere, which is a round ball centered at the origin. Because its radius is much greater than the expected support of the charge (Alice's planet), $\langle Q \rangle$ will not depend on its precise choice. Thus we can allow for a time-dependent radius, for example as
\begin{equation}
r(t) = r_B + \alpha (t-t_B)~. \label{eq:linR}
\end{equation} 
Physically, this corresponds to the freedom to let Bob's Dyson sphere expand or contract during the measurement.\footnote{One might worry that $r(t)$ is negative for $t<t_B-\frac{r_B}{\alpha}.$ However, since this happens only at the tail of the weight function $w(t)$ (Eq.~(\ref{eq:w})), it does not affect our results. For example, the choice $r(t)=r_B (1-\alpha  \tanh (r_B))+\alpha  t \tanh (t),$ which has the same behavior as Eq.~(\ref{eq:linR}) at large $t$ and is nowhere negative, leads to the same asymptotic behavior.} This turns out to give Bob more freedom to suppress fluctuations, but nevertheless we will find that they diverge.

We are interested in the limit as Bob's radius goes to infinity along a lightcone, $r_B=t_B+u_B\to\infty$, so that $Q$ becomes the Bondi charge. By an overall time shift, we may set the fixed retarded time of Bob's measurement to zero, $u_B=0$. We can then fix the retarded time duration of Bob's measurement, as the interval $-\frac{\delta u}{2}<u<\frac{\delta u}{2}.$ That is, the weight function (\ref{eq:w}) should have support when Bob's world tube (\ref{eq:linR}) lies in this interval, but not outside it. To this end we choose
\begin{equation}
\delta t=\frac{\delta u}{1-\alpha}. \label{eq:dtdu}
\end{equation}
Note that the proper time duration of Bob's measurement is then given by  
\begin{equation}
\delta\tau=\delta u \sqrt{\frac{1+\alpha}{1-\alpha}}.
\label{eq-proper}
\end{equation}
Intuitively, we might expect that fluctuations will be more suppressed for greater $\delta\tau$, i.e., for Bob's sphere expanding at great velocity, $\alpha\to 1$. However, as we shall see this is not sufficient to control the fluctuations as $r_B\to\infty$.

To evaluate $\langle Q^2 \rangle$, we now write it as
\begin{equation}
\langle Q^2 \rangle\! = \!\int \!\! d^dx \!\! \int\!\! d^d\! \Delta\, 
w(x^0) w(y^0) \theta (\vec x) \theta (\vec x - \vec \Delta)\langle j^0(0) j^0(\Delta) \rangle, \label{eq:Qsq1}
\end{equation}
where $\theta=1$ inside the volume $V$ and $\theta=0$ outside. 

Here we summarize how this calculation goes. More details can be found in the Appendix. The integral over $d^{3} \vec x$ yields the volume of the intersection of two balls separated by $|\vec \Delta|$. By spherical symmetry, the integral over $d^{3} \vec \Delta$ reduces to a one-dimensional integral which we evaluate. We subsequently perform the $dx^0$ and $d\Delta^0$ integrations using contour methods. Here one has to be careful to choose a contour that properly avoids branch cuts. This yields an expression for $\langle Q^2 \rangle$ as a function of $r_B, \delta t$, and thus via Eq.~(\ref{eq:dtdu}), of $r_B,\delta u,\alpha$.

\begin{eqnarray}
 \langle Q^2 \rangle &=& -\kappa \left(\pi^2 \frac{(1-\alpha)^3 r_B^2}{3 (\alpha +1)
   {{\delta u}}^2}+\frac{\pi^2}{6} \log \left(\frac{4(1-\alpha)^3 r_B^2}{(\alpha +1)
   {{\delta u}}^2}\right) \right) \nonumber \\ 
   &-& \frac{\kappa \pi^2}{12(\alpha^2-1)} + O\left(r_B^{-1}\right)  \label{eq:Qsqans}
\end{eqnarray}

We can now take the limit $r_B\to\infty.$ For $\alpha=0$, we find an expected area law divergence. For other choices of $\alpha,$ it is possible to have $\langle Q^2 \rangle$ diverge slower than that. To accomplish the goal of making $\langle Q^2 \rangle$ grow as slow as possible with $r_B,$ the optimal choice of $\alpha$ satisfies  

\begin{equation}
1-\alpha^{\rm opt} \propto \sqrt{\frac{\delta u}{r_B}}, \label{eq:alphaopt}
\end{equation}

No choice of $\alpha$ can make $\langle Q^2\rangle $ diverge slower than that, and in particular, no choice of $\alpha$ can make the charge fluctuations finite when $r_B \rightarrow \infty.$ For the optimal choice above, the divergence goes as the fourth-root of the area,
\begin{equation}
\langle Q^2 \rangle^{\rm opt} \sim \sqrt{\frac{r_B}{\delta u}}~.
\label{eq-qflucopt}
\end{equation}

The results above are for four dimensional Minkowski space, but the same analysis can be performed in any dimension (though we have only been able to get analytic results in even dimensions). Here we quote the results in two\footnote{Since QED is confining in 2D, one cannot give the 2D result the same interpretation as in higher dimensions.} and six dimensions:

\begin{eqnarray}
\langle Q^2 \rangle_{D=2} &\propto& \log \left(\frac{\left({\delta u}^2+(1-\alpha)^4 {r_B}^2\right)^2}{\left(1-\alpha
   ^2\right)^2 {\delta u}^4}\right) \\
   \langle Q^2 \rangle_{D=6}&\propto& \frac{(1-\alpha )^6 {r_B}^4}{(\alpha +1)^2
   {\delta u}^4}+O\left({r_B}^2\right)
\end{eqnarray}

We see that for constant $\alpha,$ we always get an area law $\langle Q^2 \rangle_D \sim\left(\frac{r_B}{\delta u}\right)^{D-2}.$ The optimal choice of $\alpha$ is always given by Eq.~(\ref{eq:alphaopt}) for any $D$; this yields \begin{equation}
\langle Q^2 \rangle^{\rm opt}_D \sim r_B^{(D-2)/4} \sim \delta\tau^{D-2}~.
\end{equation}

This divergence thwarts Bob\rq{}s plans of measuring the charge and thus prevents him from receiving Alice\rq{}s message. Since no information is transmitted, the apparent paradox described in the previous section is resolved.

\section{Bondi mass \label{gravity}}

In the previous section we showed that, due to quantum fluctuations, the Bondi electric charge cannot be measured in a finite interval of $\scri^+.$ Here we repeat this analysis, but for the Bondi mass. For concreteness, we consider a massless scalar field non-minimally coupled to gravity. However, since the two point function of $T_{00}$ is completely fixed (up to a multiplicative factor) in any scale-invariant theory with a stress-tensor, our conclusions apply equally well to spinors, gauge fields, and interacting fixed points.

The action and stress-energy tensor for a non-minimally coupled scalar are given by 
\begin{equation}
S=-\frac{1}{2} \int d^4 \sqrt{-g}\left(D_\mu \phi D^\mu \phi + \xi R \phi^2 \right),
\end{equation}
and
\begin{eqnarray}
T_{\alpha \beta} &=& (1-2\xi) D_\alpha \phi D_\beta \phi + \left(2\xi - \frac{1}{2}\right) D_\mu \phi  D^\mu \phi g_{\alpha \beta} \nonumber \\
&+& 2 \xi g_{\alpha \beta} \phi D^2 \phi - 2 \xi \phi D_\alpha D_\beta \phi.
\end{eqnarray}

Using this stress-energy tensor and $\langle \phi(0) \phi(\Delta)\rangle = \frac{1}{\Delta^2},$ we get
\begin{equation}
\langle T_{00}(x)T_{00}(y)\rangle = 8 \left(30 \xi ^2-10 \xi +1\right) \frac{3 {\vec \Delta}^4+10 \Delta_0^2{\vec \Delta}^2+3 \Delta_0^4}{\left(\Delta_0^2-{\vec \Delta}^2\right)^6}. \label{eq:TT}
\end{equation}
Using the same smearing as in the previous section, we can now calculate the fluctuations of the energy,
\begin{equation}
\langle M^2 \rangle =  \!\int \!\! d^4x \!\! \int\!\! d^4\! \Delta\, 
w(x^0) w(y^0) \theta (\vec x) \theta (\vec x - \vec \Delta) \langle T_{00}(x)T_{00}(y)\rangle, \label{eq:Esq}
\end{equation}
by performing the same integrals as in the QED case, the details of which are relegated to the Appendix. 

As in the $U(1)$ case, we choose to evaluate the operator and its fluctuations as a volume integral, not a surface integral. This is now more subtle, because strictly the Bondi mass is defined {\em only} as a surface integral over a family of topological 2-spheres $\left \{ S_{\alpha} \right \}$ that approach a cut $S$ of null infinity \cite{Wald:1984rg}:
\begin{equation}
M = -\lim_{S_{\alpha}\rightarrow S}\frac{1}{8\pi}\int_{S_{\alpha}}\varepsilon_{abcd}\nabla^c \zeta^d
\label{eq-wald}
\end{equation}
where $\zeta^a$ is an asymptotic time translation Killing vector field. Here we work in a perturbative limit, where backreaction in the bulk is small. Then an approximate Gauss law still holds, and the Bondi mass can also computed as a volume integral
\begin{equation}
M = \int_{\tilde{\Sigma}} d^3x \ T_{00} 
\end{equation}
over the portion $\tilde{\Sigma}$ of a Cauchy surface $\Sigma$ enclosed by $S$. Moreover, we can reach arbitrarily large $M$ even in the perturbative regime, by considering matter of low density spread over a large region. Hence we expect that our result for the fluctuations of $M$ will be general.

We find 
\begin{eqnarray}
 \langle M^2 \rangle &=&  8 \left(30 \xi ^2-10 \xi +1\right) \pi ^2 \left(\alpha ^2 {\delta u}^2+4 (1-\alpha)^4
   {r_B}^2\right)^3 \nonumber \\&\times& \left((1-\alpha)^4 \left(3 \alpha ^2+1\right)
   {r_B}^2-\left(\alpha ^2-5\right) \frac{{\delta u}^2}{4}\right) \nonumber \\ &\times & \left(15 (1-\alpha)
   (\alpha +1)^3 {\delta u}^4 \left({\delta u}^2+4 (1-\alpha)^4
   {r_B}^2\right)^3\right)^{-1} \nonumber \\ \label{eq:M2gen}
\end{eqnarray}
For $\alpha=0$ this gives 
 \begin{equation}
 \langle M^2 \rangle = 8 \left(30 \xi ^2-10 \xi +1\right)  \frac{16 \pi ^2 r_B^6 \left(5 {\delta u}^2+4 r_B^2\right)}{15 {\delta u}^4
   \left({\delta u}^2+4 r_B^2\right)^3}. \label{eq:M2a0}
 \end{equation}
Once again, it is possible to tame this divergence by a better choice of $\alpha.$ The optimal value remains $\alpha^{\rm opt} \propto 1-\left(\frac{r_B}{\delta u}\right)^{-1/2},$ which gives
\begin{equation}
\langle M^2 \rangle^{\rm opt} = \frac{\left(30 \xi ^2-10 \xi +1\right) 2^{5/2} \pi ^2}{30 \delta u^{5/2}}
  \sqrt{r_B}+O\left(\frac{1}{r_B^{1/2}}\right). \label{eq:M2aopt}
\end{equation}

We therefore see that the Bondi energy also has unbounded fluctuations as we approach finite intervals of null infinity.  
 
\section{Discussion}

We argued that entropy bounds preclude gauge charges from being well-defined quantum observables on cuts or finite intervals of $\scri^+$. We confirmed this by showing that unbounded fluctuations preclude a measurement of the electric charge or the Bondi mass, in finite time at arbitrarily large radius.\footnote{The study of fluctuation of electric charge (in finite regions) dates back to the early days of QED  (see e.g. \cite{PhysRev.47.144} and \cite{PhysRev.78.794}).} 

It is important to emphasize the quantum nature of these results. Both $M$ and $Q$ are good classical observables near a cut of ${\scri}^+$. This follows directly from Eq.~(\ref{eq-q}), and from the analogous surface integral for the Bondi mass, Eq.~(\ref{eq-wald}). Both expressions are gauge-invariant and require no data extrinsic to the near-cut region ${\cal R}$ for their evaluation. This constrasts with certain other quantities appearing in the Bondi metric expansion, Eq.~(\ref{Bondi Expansion}), which are prohibited by the equivalence principle from being observable already at the classical level \cite{BouHal16,BouChaHal}.

Let us try to gain some intuition for the divergence of $\langle Q^2 \rangle$ and $\langle M^2 \rangle$ that we found.
To understand the physical origin of the fluctuations, suppose, for simplicity, that Bob remains at fixed radius throughout his measurement, so that $\alpha=0$ and $\delta u=\delta t=\delta \tau$.  Consider $Q$ as a surface integral over $E_r$, rather than a volume integral. An observation restricted to a finite time interval leads to approximately thermal quantum noise of characteristic energy $1/\delta \tau$. This noise arises in the region causally accessible to the observer; here, this would be a shell of width $\delta t$ around the sphere $r_B$. Since $r_B\gg \delta \tau$, there will be a large number $N\sim r_B^2/\delta \tau^2$ of ``cells'' just inside and outside of Bob's sphere. Each cell contains $O(1)$ quanta of any massless field the detectors couple to, which includes the charges. This contributes to $E_r$ an additional field strength of order $1/\delta \tau^2$ and random sign. The contribution to $Q$ from one cell, in Eq.~(\ref{eq-q}), is thus of order $\pm 1$. The fluctuations in different cells are uncorrelated, so the total fluctuation of $Q$ is given by $\langle Q^2 \rangle^{1/2}\sim \sqrt{N}\sim r_B/\delta \tau$. This agrees with Eq.~(\ref{eq:Qsqans}) for this special case, $\alpha=0$.\footnote{It would be nice to extend this heuristic argument to the optimal case, when Bob is expanding outward during the measurement according to Eq.~(\ref{eq:alphaopt}). But using Eq.~(\ref{eq-proper}), the above argument would appear to imply $\langle Q^2 \rangle\sim r_B^2/\delta \tau^2\sim (r_B/\delta u)^{3/2}$, in conflict with Eq.~(\ref{eq-qflucopt}).}

Note that neither infrared nor ultraviolet physics alone can explain the divergent fluctuations of $Q$ and $M$. Rather, they arise from a combination of both. The fixed duration $\delta u$ of Bob's measurement sets a characteristic ``ultraviolet'' energy scale for the fluctuations. The infrared effect comes from taking the limit as $r_B\to \infty$, which creates an ever larger region over which those fluctuations can contribute. 

Our work lends some insight on the structure of operator algebras of gauge theories and gravity when quantizing at $\scri^+$.  We emphasize that the paradox noted in Section \ref{paradox} would arise for any quantity associated to a subset of ${\scri}^+$ that is not tied to energy flux arriving in that subset. For example, the BMS group at $\scri^+$ yields an infinite set of supertranslation charges \cite{Barnich:2011mi}, which essentially correspond to the Bondi mass aspect (whose integral yields the Bondi mass)~\cite{Ashtekar:1981sf,Str13, StrZhi14,Strominger:2017zoo}. We thus find that these supertranslation charges are not observable
in a neighborhood of any cut of $\scri^+$ in the quantum theory\footnote{We established that a certain operator $\hat O$ does not belong to the algebra of observables by showing that $\langle \hat O^2 \rangle= \infty$. This is not a perfect criterion, since there are contrived examples of observables in quantum mechanics with $\langle \hat O^2 \rangle=\infty$ but well-defined spectrum. However, we do expect all \emph{reasonable} operators to have finite fluctuations.}. 

The absence of such observables also has potential significance for understanding the holographic principle.  There has been considerable interest in trying to construct a holographic theory dual to asymptotically flat spacetimes (see \cite{Cheung:2016iub, Kapec:2016jld, Pasterski:2016qvg} for recent examples).  By analogy to AdS/CFT, one expects that such a putative holographic dual should be defined on the conformal boundary of the spacetime, and that limits of bulk observables that are defined as they approach $\scri^+$ should correspond to local operators in the putative boundary theory.  Since we have shown that conserved charges are not in fact well-defined operators on any finite portion of $\scri^+$, we expect that no such operators should exist in a dual boundary theory either. 
 
\paragraph*{Acknowledgments} It is a pleasure to thank Arvin Shahbazi Moghaddam for discussions. We also thank Juan Maldacena, David Simmons-Duffin and Anthony Speranza for discussions about the first version of this manuscript.
This work was supported in part by the Berkeley Center for Theoretical Physics, by the The Institute for Advanced Study, by the National Science Foundation (award numbers 1521446, 1316783 and 1314311), by FQXi, by the S. Raymond and Beverly Sackler Foundation Fund, by SITP, by the ``It from Qubit'' Collaboration (Simons Foundation), and by the US Department of Energy under contract DE-AC02-05CH11231. 

\begin{widetext}
\appendix 

\section{Calculation of $\langle Q^2 \rangle$ and $\langle M^2 \rangle$}
In this appendix we describe in more detail the calculations of $\langle Q^2 \rangle$ described in section \ref{QED} and of $\langle M^2 \rangle$ described in section \ref{gravity}. 

For the QED calculation, we start with eq.(\ref{eq:Qsq1}). Inserting the expression for the current 2-point function, Eq.~(\ref{eq-2j}), and evaluating the $d^3 \vec x$ integral, we get 
\begin{equation}
\langle Q^2 \rangle = \kappa \int dx^0 d \Delta^0 4 \pi^2 d \Delta \frac{\left(\Delta ^2+(\Delta^0)^2\right) \Vol(r_1(t),r_2(t),\Delta)}{ \left(\Delta^2-(\Delta^0)^2\right)^2} w(x^0) w(x^0-\Delta^0), \label{eq:2DRofts}
\end{equation}
Note that here $\Delta=|\vec \Delta|,$ whereas in the main text $\Delta$ denoted a four-vector.  

The radii, as functions of time, are specified by 
\begin{eqnarray}
r_1(t)=r_B + \alpha (t-t_B), \nonumber \\ 
r_2(t)=r_B + \alpha (t-\Delta^0-t_B),  \label{eq:r12}
\end{eqnarray}

$w(t)$ is given in Eq.(\ref{eq:w}) and $\Vol(r_1,r_2, \Delta)$ is the volume of the intersection of two spheres of radii $r_1$ and $r_2$ whose centers are separated a distance $\Delta.$ Explicitly, the volume formula is 
\begin{equation}
\Vol(r_1,r_2, \Delta) =
\frac{\pi  (-\Delta +{r_1}+{r_2})^2 \left(\Delta ^2-3 \left({r_1}^2+{r_2}^2\right)+2 \Delta  ({r_1}+{r_2})+6 {r_1}
   {r_2}\right)}{12 \Delta },
\end{equation}
for  $|r_1-r_2| \leq \Delta \leq r_1+r_2.$ For $\Delta>r_1+r_2,$ the spheres do not intersect, and so $\Vol(r_1,r_2, \Delta) =0.$ For $\Delta <| r_1-r_2|,$ one ball is inside the other and so $\Vol(r_1,r_2, \Delta) =\frac{4}{3} \pi \min(r_1,r_2)^3.$ Evaluating the $\Delta$ integral in Eq.~(\ref{eq:2DRofts}), we get

\begin{equation}
\langle Q^2 \rangle = \frac{16 \pi^6 \kappa}{15} \int d x^0 d \Delta^0 \frac{{r_1}^3 {r_2}^3 (-5 {(\Delta^0)}^4+2 {(\Delta^0)}^2
   \left({r_1}^2+{r_2}^2\right)+3
   \left({r_1}^2-{r_2}^2\right)^2)}{(\Delta^0+{r_1}-{r_2})^3 (\Delta^0+{r_1}+{r_2})^3   (-\Delta^0+{r_1}-{r_2})^3 (-\Delta^0+{r_1}+{r_2})^3}
   \frac{ \delta t^2 / \pi^2}{\left(\delta t^2+{x^0}^2\right) \left(\delta t^2+(x^0-{(\Delta^0)}^2)\right)}. \label{eq:Qsq2}
\end{equation}

We now choose contours to evaluate, in turn, the $\Delta^0$ and the $x^0$ integrals. Keep in mind that at this step the expressions for $r_1$ and $r_2,$ Eq.(\ref{eq:r12}), need to be explicitly inserted. Seen as a function on the complex $\Delta^0$ plane, the integrand in Eq.~(\ref{eq:Qsq2}) has four branch points, all on the real axis, and two simple poles, at $\Delta^0=x^0 \pm i \delta t.$ We choose a contour that goes along the real axis, with infinitesimal deformations around the branch points to avoid them, and then close along a semi-circle on the upper half-plane (See Figure \ref{fig:cint}). This contour picks up a residue at $\Delta^0=x^0 + i \delta t,$ thus yielding

\begin{eqnarray}
\langle Q^2 \rangle &=& \int d x^0 \frac{- \pi^4 {{\delta t}} \kappa \left(\frac{8 ((\alpha -1) r_B+i {{\delta t}} \alpha ) ((\alpha -1) r_B-\alpha  {x^0})
   \left(-\left(\alpha ^2-1\right) {{\delta t}}^2-2 i {{\delta t}} ({x^0}-(\alpha -1) \alpha 
   r_B)+(\alpha -1) \left(2 (\alpha -1) r_B^2-2 \alpha  r_B {x^0}+(\alpha +1)
   (x^0)^2\right)\right)}{(\alpha -1) (\alpha +1) ({{\delta t}} (\alpha +1)-i (\alpha -1) (2
   r_B-{x^0})) ({{\delta t}} (\alpha -1)-i (2 (\alpha -1) r_B-(\alpha +1) {x^0}))}\right)}{12
   \pi ^3 ({{\delta t}}-i {x^0})^3 ({{\delta t}}+i {x^0})}\nonumber\\
   &-&\frac{({{\delta t}}-i {x^0})^2 \log
   \left(\frac{\left({{\delta t}}^2+(-2 (\alpha -1) r_B+\alpha  ({x^0}-i {{\delta t}}))^2-2 i {{\delta t}}
   {x^0}-(x^0)^2\right)^2}{\left(\alpha ^2-1\right)^2 ({x^0}+i {{\delta t}})^4}\right)}{12
   \pi ^3 ({{\delta t}}-i {x^0})^3 ({{\delta t}}+i {x^0})}
\end{eqnarray}

Looking at the integrand above as a function of $x^0$ on the complex plane we see that the branch points, in the limit of interest ($\alpha \rightarrow 1_-$), do not lie above the real line. Thus, the same contour prescription can be used to evaluate the $x^0$ integral, which now picks up a residue only at the simple pole at $x^0= i \delta t.$ Doing so, and using Eq.(\ref{eq:dtdu}) to replace $\delta t$ to $\delta u$, gives

\begin{equation}
\kappa \pi^2\frac{-\alpha ^2 \left(\alpha ^2-2\right) {\delta u}^4+8 (\alpha -1)^4
   {\delta u}^2 {r_B}^2-\left(1-\alpha ^2\right) {\delta u}^2
   \left({\delta u}^2+4(\alpha -1)^4 {r_B}^2\right) \log
   \left(\frac{\left({\delta u}^2+4 (\alpha -1)^4 {r_B}^2\right)^2}{\left(\alpha
   ^2-1\right)^2 {\delta u}^4}\right)+16 (\alpha -1)^8 {r_B}^4}{12 (\alpha -1)
   (\alpha +1) {\delta u}^2 \left({\delta u}^2+4 (\alpha -1)^4 {r_B}^2\right)}.
\end{equation}

\begin{figure}[h] 
\includegraphics[scale=0.13]{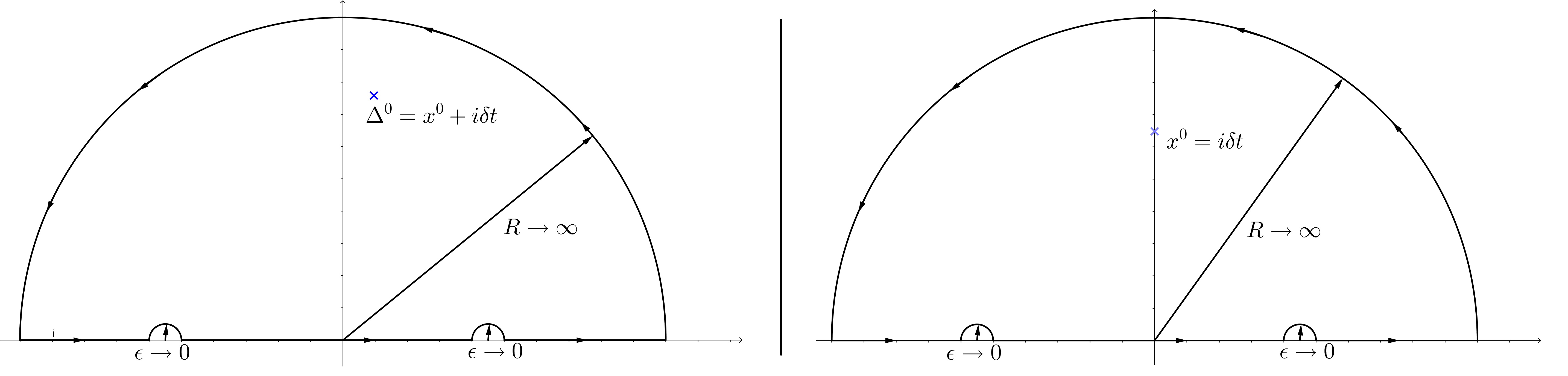}
\caption{In the $\Delta^0$ integral (left diagram), the contour avoids the branch points on the real axis and picks up a residue at the simple pole at $\Delta^0=x^0+i \delta t.$ In the $x^0$ integral (right diagram), a similar contour is used. It now picks up a residue at the simple pole at $x^0=i \delta t.$}\label{fig:cint}
\end{figure}

The series expansion of this at large $r_B$ gives the result in Eq.~(\ref{eq:Qsqans}). We have also checked that this agrees with the result of numerically integrating Eq.~(\ref{eq:Qsq2}).

The calculation of energy fluctuation in the null infinity limit parallels the calculation above. For concreteness, let\rq{}s consider a scalar field, and take as our starting point Eq.~(\ref{eq:Esq}). Inserting Eq.~(\ref{eq:TT}), and evaluating the $d^3 \vec x$ integral, we get

\begin{equation}
\langle M^2 \rangle = \!\int \!\! d x^0 d \Delta^0 4 \pi^2 d \Delta \Vol(r_1,r_2, \Delta)  8 \left(30 \xi ^2-10 \xi +1\right) \frac{3 {\vec \Delta}^4+10 \Delta_0^2{\vec \Delta}^2+3 \Delta_0^4}{\left(\Delta_0^2-{\vec \Delta}^2\right)^6} w(x^0) w(\Delta^0).
\end{equation}

Evaluating the $\Delta$ integral gives 

\begin{equation}
\langle M^2 \rangle = -\frac{ \left(30 \xi ^2-10 \xi +1\right) 128 {\delta t}^2 {r_1}^3 {r_2}^3 \left(-5 (\Delta^0)^4+2
   (\Delta^0)^2 \left({r_1}^2+{r_2}^2\right)+3
   \left({r_1}^2-{r_2}^2\right)^2\right)}{15 \left({\delta t}^2+(x^0)^2\right)
   \left({\delta t}^2+(x^0-{\Delta^0})^2\right) (-\Delta^0+{r_1}-{r_2})^3 (-{\Delta^0}+{r_1}+{r_2})^3
   ({\Delta^0}+{r_1}-{r_2})^3 (\Delta^0+{r_1}+{r_2})^3}
\end{equation}

Following the same contour prescription as before (see Figure \ref{fig:cint}), the $\Delta^0$ integral picks up the residue at $\Delta^0=t+ia$ and evaluates to

\begin{eqnarray}
&& \langle M^2 \rangle =\frac{128  \left(30 \xi ^2-10 \xi +1\right)  \pi  {\delta t} ((\alpha -1) r_B+i {\delta t} \alpha )^3 (-\alpha  r_B+r_B+\alpha  x^0)^3}{15 (\alpha
   -1)^3 (\alpha +1)^3 ({\delta t}-i x^0)^5 ({\delta t}+i x^0) ({\delta t} (\alpha +1)-i (\alpha -1) (2
   r_B-x^0))^3 ({\delta t} (\alpha -1)-i (2 (\alpha -1) r_B-(\alpha +1) x^0))^3} \nonumber \\
   &\times&  [ \left(3 \alpha ^4+2 \alpha ^2-5\right) {\delta t}^2+2 i {\delta t} \left(\left(3
   \alpha ^4+5\right) x^0-2 \alpha  \left(3 \alpha ^3-3 \alpha ^2+\alpha
   -1\right) r_B\right)  \nonumber \\
  &-&(\alpha -1) \left(4 \left(3 \alpha ^3-3 \alpha
   ^2+\alpha -1\right) r_B^2-4 \left(3 \alpha ^3+\alpha \right) r_B x^0+\left(3
   \alpha ^3+3 \alpha ^2+5 \alpha +5\right) (x^0)^2\right) ]
\end{eqnarray}

A similar contour can be used for the $x^0$ integral now, which picks up a residue at $t=i a,$ and gives the answer in Eq.~(\ref{eq:M2gen}).

\end{widetext}

\bibliographystyle{utcaps}
\bibliography{all}

\end{document}